\documentclass[a4paper,notitlepage]{article}
\usepackage{graphicx}
\usepackage{amsmath}
\usepackage{a4}
\usepackage{amsfonts}
\usepackage{amssymb}
\begin{document}

\title{Ripples in Tapped or Blown Powder}
\author{Jacques Duran\\LMDH- UMR\ 7603\ CNRS- Universit\'{e} P.\ et M.\ Curie\\4 place Jussieu, 75252 Paris Cedex 05\\email address : jd@ccr.jussieu.fr}
\maketitle

%

%TCIMACRO{\TeXButton{abstract}{\abstract{
%We observe ripples forming on the surface of a granular powder in a
%container submitted from below to a series of brief and distinct shocks.
%After a few taps, the pattern turns out to be stable against any further
%shock of the same amplitude.\ We find that the wavelength of the pattern is
%proportional to the amplitude of the shocks.\ Starting from considerations
%involving air flow through the porous granulate and avalanche properties, we
%build up a semi-quantitative model which satisfactorily fits the set of
%experimental observations of either tapped or blown powder.}}}%
%BeginExpansion
\abstract{
We observe ripples forming on the surface of a granular powder in a
container submitted from below to a series of brief and distinct shocks.
After a few taps, the pattern turns out to be stable against any further
shock of the same amplitude.\ We find that the wavelength of the pattern is
proportional to the amplitude of the shocks.\ Starting from considerations
involving air flow through the porous granulate and avalanche properties, we
build up a semi-quantitative model which satisfactorily fits the set of
experimental observations of either tapped or blown powder.}%
%EndExpansion

\bigskip

In the recent years, there has been a great deal of interest in the response
of granular materials\ to various kinds of external perturbations.\ Up to now,
the vast majority of the experimental, theoretical and simulated works have
dealt with model granular solids in the sense\cite{brown70} that the particles
were supposed to be large enough (i.e.\ typically larger than $100\mu m)$ to
avoid significant interaction with the surrounding fluids. In reverse and
rather paradoxically, the understanding of the behavior of fine powders has
received much less attention although it is universally recognized as the
keystone of an increasing number of high-tech industrial processes. In this
spirit, a few recent attempts were made towards the analysis of the behavior
of fine cohesive (e.g. \cite{castellanos99}) or non-cohesive powders
(typically in the range from $1\mu m$ to less than $30\mu m)$ or of larger
particles submitted to excessive windy conditions, involving saltation such as
in desert dunes.

Among others, our present knowledge about instability of layers of granular
solids under vertical vibrations is currently firmly established.\ As a matter
of fact, the original paper of Melo \emph{et al.}\cite{melo94} has stimulated
a series of subsequent papers\cite{umbanhowar96}\cite{melo95} \cite{clement96}%
\cite{tsimring97} dealing with several facets of the same problem.\ In short,
all these works converged toward a description of the observed dynamic
behaviors in terms of Faraday instability in liquids combined with the
specific dynamics of the inelastic bouncing ball.

In a less recent past, the vibrational heaping of a sand pile has motivated a
debate\cite{fauve89}\cite{evesque89} \cite{clement92} dealing with the
influence of the air interaction in the phenomenology of sand heaping.

The present work\ (which duplicates into a small scale laboratory experiment,
a real industrial device used to empty powder carrying tankers) reports the
observation and analysis of a novel form of surface instability of a thin
layer (thickness typically $10mm)$ of a fine powder ( particle size $10\mu m)$
submitted to a series of vertical shocks from below the container.\ A
relatively simple experiment exhibits two essential features which give a
trend to a plausible semi-quantitative theoretical explanation of the
process.\ First, it is observed that, after a few taps, the surface displays a
regularly corrugated pattern made of a succession of jointed heaps sitting at
the natural avalanche angle.\ The crucial point here is that any further taps
do not induce any significant change in the pattern which thus can be
considered as a steady state with respect to further vertical shocks.\ Second,
and this is a clue to the understanding of the process, the characteristic
wavelength of the pattern is found to be directly proportional to the
amplitude of the taps.

Our simple analytical model considers the powder-air interaction.\ It involves
two basic features of the fine granular material : Firstly, the Darcy's law
for modelling the air flux through the porous cake of granulate.\ Secondly,
the maximum stability angle of a granulate before generating avalanches.\ In
brief, it is shown that a corrugated surface stands as a more stable state
than a horizontal flat surface with respect to air blow from below, because it
is easier to eject a particle from a flat surface than from an inclined
surface sitting at the avalanche angle. Additionally, the apices of the hills
created by air blow are seen to be unstable as compared to both sides of the
hills. A couple of further experiments provide a clear view of this last feature.

\begin{figure}[h]
\begin{center}
\includegraphics[
height=3.3053in,
width=1.4589in
]{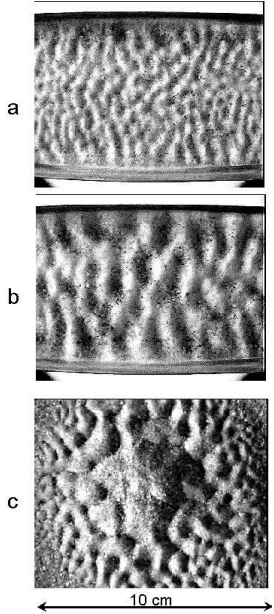}
\end{center}
\caption{Three bird-eye views of the corrugated surface observed after twenty
shocks of constant amplitudes onto the underpart of the containers.\ Snapshots
a and b corresponds to the cylindrical container and are obtained at two
different shock amplitudes, larger in b than in a. Measurements are performed
in the median part of the pattern. Snapshot c is obtained in a rectangular
metallic box (size 20x40cm$^{2})$ containing a layer of fine sand beach,
tapped under the central part. In this latter case, the pattern reproduces the
transient deformation of the underlying metallic sheet.}%
\label{Fig1}%
\end{figure}

Several basic characteristic features of the surface instability of a tapped
powder layer can be readily observed starting from a simple table-top
experiment : We use a cylindrical transparent tube made of leucite or
glass.\ The dimensions of the tube are unimportant.\ In a typical experiment,
this tube can be about 20$cm$ long and $2.5cm$ in diameter.\ We half fill the
tube with fine powder (e.g. glass beads, diameter $10\mu m)$.\ We keep the
tube horizontal and rigidly fixed at both ends, with the powder initially set
flat and horizontal thus giving a granulate thickness of about $12.5$mm in the
center.\ We take a heavy metallic or plastic rod and knock gently and
repeatedly at a very low pace and at a constant intensity onto the center of
the tube from below, applying vertically as brief taps as possible,
i.e.\ letting the rod rebound appreciably after each separate shock. After a
few taps (about ten to twenty), the surface, initially flat, smooth and
horizontal, turns out to exhibit ripples similar to those reported in Fig.
\ref{Fig1}a and \ref{Fig1}b. Now, tapping $\emph{more}$ \emph{energetically}
but still keeping the intensity as constant as possible from one tap to the
next, induces a pattern where the mean distance between two successive ripples
increase significantly. Furthermore and under energetic tapping, a careful
observation of the surface shows that, at every tap, a limited number of
particles may be ejected upwards starting both from the apices of the hills
and from the small plateaux which happen occasionally between imperfectly
jointed hills. Using a flat rectangular box as a container instead of a
cylindrical tube, and tapping under the center of the box gives rise to
surface patterns similar to Fig.\ \ref{Fig1}c. Note that similar results can
be obtained using rather large and hollows i.e. light particles. Excessive
wetness prevents the observation of these surface patterns.

\begin{figure}[h]
\begin{center}
\includegraphics[
height=3.3918in,
width=4.4858in
]{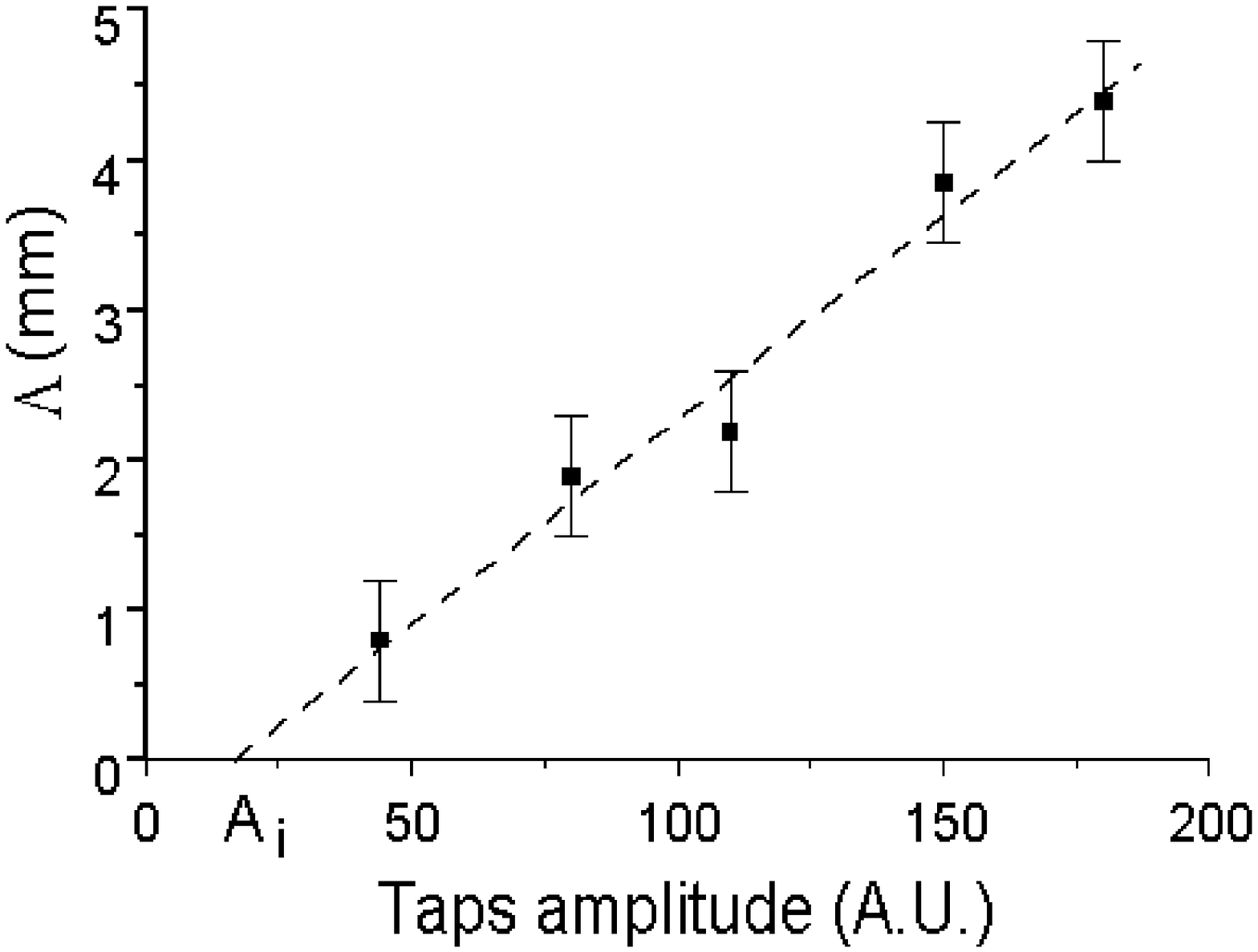}
\end{center}
\caption{Characteristic wavelength of the pattern in millimeter versus taps
amplitude measured as the signal delivered by the microphone in tens of
millivolts. The dotted line shows a linear fit of the experimental results.}%
\label{Fig2}%
\end{figure}

Definitely more reliable information has been obtained in the course of our
experiments, using a more sophisticated device. We set a CCD (charge coupled
device) camera above the tube in order to record and process the successive
patterns obtained during the experiments.\ Secondly, we used a magnetically
driven tapping device and a microphone stuck on the tube in order to monitor
the amplitude of the taps applied on the sample. Typical experimental results
are reported in Fig.\ \ref{Fig2}

Considering the peculiar geometry of the pattern made of jointed heaps sitting
approximately at avalanche angle, we note that the characteristic wavelength
$\Lambda$ is determined by the height of the pattern.\ We call $\ h_{T}$ the
altitude (starting from the bottom of the container) of the apices of the
corrugated surface, $h_{B}$ the altitude of the valleys of the corrugated
surface and $h_{i}$ the altitude of the initially horizontal surface of the
granular layer. $\theta$ is the avalanche angle of the powder, which is about
$30%
%TCIMACRO{\UNICODE{0xb0}}%
%BeginExpansion
{{}^\circ}%
%EndExpansion
$ in our glass beads powder.

The wavelength $\Lambda$ is given by%

\begin{align}
\Lambda &  =2(h_{T}-h_{B})\cot\theta\\
\text{where \ }h_{T}+h_{B}  &  =2h_{i}%
\end{align}

We first start from the first part of an assertion initially put forth by
M.\ Faraday\cite{faraday31} who stated that in the course of vertical
vibration of a granular powder ''It forms a partial vacuum into which the air
round the heap enters with more readiness than the heap itself,...'' but
instead of following immediately the second part of his statement (stating
that it results in sand heaping by carrying sand at the bottom edge with it),
we rather consider here that, during the fall following the upward launching
of the sand by the taps, the trapped air is forced upwards through the porous
medium.\ Then, supposing for the sake of simplicity that the air flux through
the medium obeys a Poiseuille flow, the Darcy's law holds true. It gives the
velocity $v$ of a fluid flux emerging from the porous medium whose
permeability is $K$ and where the air is pushed upwards by a uniform pressure
difference $\Delta P$ acting over a thickness $h$ of the granulate as%

\begin{equation}
v_{h}=K\frac{\Delta P}{h}%
\end{equation}

In the course of our experiments $\Delta P$ is proportional to the tap
amplitude\ applied on the underpart of the tube $A$ so that the velocity of
the air emerging from the surface at altitude $h$ can be written as
$v_{h}=\alpha A/h$

where $\alpha$ is the coefficient of proportionality given by Darcy's law
which involves the permeability of the granular material.

A single spherical bead (diameter $D$) is lifted up by the air flow if the air
velocity is larger than a minimum velocity $v_{f}$ ( $f$ standing for free
fall velocity) given by $v_{f}=\frac{D^{2}}{18\eta}\rho g$ where $\eta$ is the
viscosity of the gas (air), $\rho$ is the volumetric density of the particles
and $g$ the gravitational acceleration.

Thus a particle deposited on the surface at altitude $h$ can be expelled from
an horizontal surface of the layer if the outgoing air velocity is $v_{h}\geq
v_{f}$ .\ We find that for glass particles of diameter=$10$ $\mu m$, this
threshold air velocity is about $1cm/s.$ Thus the threshold quantity $\alpha
A_{i}=h_{i}v_{f}$ required for external taps able to blow up particles from
the surface of the granulate is about\ $1%
%TCIMACRO{\unit{cm}}%
%BeginExpansion
\operatorname{cm}%
%EndExpansion
^{2}/s$.

First, in view of our experimental observation, we easily realize that the air
slowing down process through the granular layer is unable to account alone for
the above described observations.\ If we imagine that the incoming air pulse
is unable to eject particles when reaching the apices of the hills, we have
$\alpha A_{T}<h_{T}v_{f}$

Calculating the ratio of the required initial velocity to induce hills of
height $h_{T}$ to the required initial velocity for the onset of the
corrugation, we get $\ A_{T}/A_{ic}=h_{T}/h_{i}$

Our experiments show that the ratio $h_{T}/h_{i}$ is only marginally larger
than $1$\ while the observed amplitude ratio is about $8$ (Fig.\ \ref{Fig2}%
).\ We conclude that another process should be taken into account to explain
the observed features.

\begin{figure}[h]
\begin{center}
\includegraphics[
height=2.7605in,
width=2.9905in
]{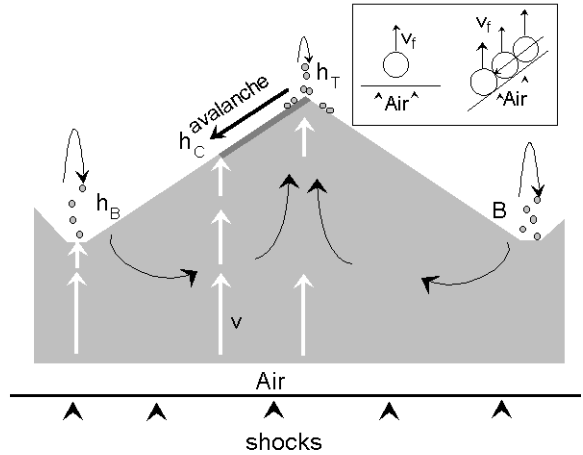}
\end{center}
\caption{Sketch of the ripples buildup showing the screening effect of the
inclined sides of the hills with respect to air blow from below. The white
arrows correspond to air trajectory while the black ones show the movement of
the particles. Insert : the balance of forces on a horizontal and on an
inclined surface.}%
\label{Fig3}%
\end{figure}

Now we can get an estimated value of the product $\alpha A$ in the course of
our experiments.\ We take profit of our observation that, in the bottom of the
valleys when a small flat surface survives or on the top of the hills when the
steady state pattern is obtained, a small number of particles are expelled at
every tap (See Fig.\ \ref{Fig3})\ 

In the course of our experiments and near the maximum tap intensity, the
upward jump of these particles is on the order of a two tenth of a
millimeter.\ There, the velocity of the emerging particles is given \ by
$\sqrt{2h_{m}g}$ $\simeq6cm/s\,\ \allowbreak$which can be considered as
approximately measuring the velocity of the emerging air flux at altitude
$h_{m}$ (about $1.2cm)$.\ Thus, we find $\alpha A\simeq7.2cm^{2}/s$ which in
view of Fig. \ref{Fig2} stands as a correct order of magnitude. Note again
that due to the fact that $h_{T}/h_{B}$ is only slightly larger than one, the
outgoing air flux is practically the same at any point on the corrugated surface.

It ensues that on both sides of the hills, the velocity of the air flux
propagating upwards is not sufficiently damped when reaching the inclined
surface to ensure the stability of the pattern. There, another process must
come into play.\ We put forward the following considerations :

On both sides of the hills, the ascending air flux meets an inclined sheet of
particles which is on the verge of avalanching$.$ Under these circumstances,
one particle subjected to the vertical incoming air flux bears a fraction of
the additional weight of the above lying particles involved in the avalanche
layer (see insert in Fig.\ \ref{Fig3})

This additional mass opposes the blowing up of the particles near the surface
and therefore stabilizes the inclined lateral surfaces against the incoming
air flow. We can build up a simplified equation for this screening effect
considering that the mass of the concerned particle is increased by a factor
$Np\sin\theta,$ $N$ being the number of the above lying particles pertaining
to a single sheet of the inclined granulate and $p$ being the unknown number
of sheets possibly involved in the avalanche process. Strictly speaking these
particles participating to the screening effect need not move, i.e.\ fall in
an avalanche process. They need only to be mobile enough to participate to the
mass balance equation involved in the process. Considering that all the
particles sitting above the considered particle sitting at altitude $h$
participate to the screening effect, we have $Np$ $\simeq\left(
h_{T}-h\right)  p/D.$ Thus the required air velocity $v_{ah}$ to eject the
considered particle sitting at altitude $h$ is given by $v_{ah}=v_{f}\left(
h_{T}-h\right)  p\sin\theta/D$

As expected, this screening effect determines an altitude $h_{C}$ under which
all the particles sitting on the inclined lateral surface cannot be expelled
by the air flux.\ This altitude $h_{C}$ is given by the equation :%

\begin{equation}
\alpha\frac{A-A_{i}}{hv_{f}}\frac{1}{\frac{h_{T}-h_{C}}{D}p\sin\theta}%
\simeq1\text{ \ \ }%
\end{equation}

In other words, the upper part of the hill ( when $h_{C}<h<h_{T})$ is unstable
whereas the lowest one ( when $h_{B}<h<h_{C})$ is stable against vertical air
blow from below.

Now, in view of several preceding observations of convective processes in
vibrational sand heaping (e.g.\cite{clement92}) experiments and in agrement
with M.\ Faraday's\cite{faraday31} second part of the statement, we note that
the steady state of the pattern should result from the balance between the
small number of expelled particles near the apices and the number of particles
which are re-injected into the bulk of the hills at every taps (see black
arrows in Fig.\ \ref{Fig3}). Starting from this observation, we conjecture
that this sort of trapping-detrapping process should be independent of the
size of the hills. Thus, we write $h_{T}-h_{C}=C(h_{T}-h_{B})$ where $C$ is
the proportion of the unstable part of the hills.\ It is an adimensional
constant independent of the height of the hills and of the amplitude of the
shocks. With this extra assumption, we get the characteristic wavelength
$\Lambda$ of the pattern which is proportional to the amplitude of the shocks
in agrement with the measurements reported in Fig.\ref{Fig2}%

\begin{equation}
\Lambda\simeq2\frac{\alpha}{C}\frac{A-A_{i}}{hv_{f}}\frac{D}{p\sin\theta}%
\cot\theta
\end{equation}

A numerical estimate using our experimental results is illustrative.\ We find
that $C=25\%$ of the hills are unstable if only one single layer of powder is
involved in the process.\ If, now, 5 layers of the superficial sheet
participate to the process as has been repeatedly observed in avalanche
experiments, we see that only $5\%$ of the upper part of the hills are
unstable against the incoming air flux.

The delicate question of the stability of the particles sitting near the
apices of the pattern has motivated a further experiment which we performed
firstly in order to prove directly the validity of our model based on
air-powder interaction and secondly to provide a visual insight into the
question of the stability of the apices of the air built pattern.

We use a millipore filter, commonly used in chemistry for filtering, in a
reverse manner. A plastic filter (pores $3\mu m)$ is placed at the bottom of a
commercial cylindrical glass vessel which allows direct observation or image
processing with a CCD\ camera. In contact with the horizontal filter and above
it, we lay a thin layer of powder (about 8mm thick). Instead of sucking up
through the filter as is usually done, we blow from below, using either brief
air pulses or a continuous air flux.

\begin{figure}[h]
\begin{center}
\includegraphics[
height=3.3062in,
width=1.6518in
]{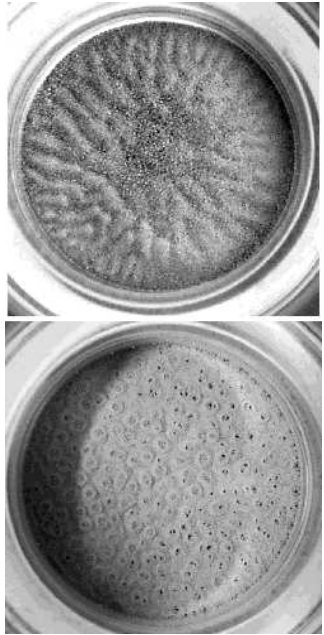}
\end{center}
\caption{The upper photograph show the ripples obtained by using fast
transient air pulses.\ The lowest photograph obtained with a continuous air
flow exhibits small craters (black points) sitting at the center of the
myriads of small fixed volcanoes.}%
\label{Fig4}%
\end{figure}

The photographs of the resulting surface corrugation are reported in Fig.
\ref{Fig4}. \ The upper snapshot shows up a surface corrugation made up of
triangular shaped ripples quite similar to the previously reported in our
tapping experiments (Fig.\ \ref{Fig1}). We again observe qualitatively that
the stronger the air pulses, the larger the characteristic wavelength of the
pattern is.\ 

The lowest snapshot corresponding to a gentle and continuous air flow going
through the powder cake is quite informative. Then the surface corrugation
exhibits a different aspect because the system has no time to relax between
separate successive perturbations as in the preceding experiments.\ This
experiment shows up a myriad of stable small volcanoes organized around small
craters (seen as black spots in the snapshot) which spew out powder particles.
This can be seen as a consistent support to our preceding picture which
involved the relative weakness of the apices of the patterns against air blow
from below.

Furthermore, we noted that the building of patterns obtained by tapping the
underpart of a granular powder is by no means a reversible effect.\ For
example, applying successively\ two series of taps at two different and
constant amplitudes gives rise to complex patterns which depend on the order
of the series.\ Doing several experiments of this type, we were able to
reproduce wrinkled volcanoes patterns or herringbone shaped structures which
strikingly remind one of their large scale counterparts in mountainous
landscapes. We postpone the description of these findings to a forthcoming paper.

I acknowledge fruitful discussions with the granular group in Jussieu, with
P.-G.\ de Gennes and with E.\ Raphael in College de France.

\end{document}